\documentclass[11pt]{article}

\usepackage{amssymb}
\usepackage{subfig}
\usepackage{color}
\usepackage{epsfig,amssymb,amsfonts,amsmath,graphicx,dsfont,cite,xfrac}
\usepackage{authblk}
\definecolor{mygray}{gray}{0.5}

\usepackage{cite}
\usepackage[colorlinks=true,linkcolor=blue,citecolor=red]{hyperref}

\newcommand{\be}{\begin{equation}}
\newcommand{\ee}{\end{equation}}
\newcommand{\bea}{\begin{eqnarray}}
\newcommand{\eea}{\end{eqnarray}}

\title{Confinement in bilayer graphene via intra- and inter-layer interactions}

\author[${}$]{Miguel Castillo-Celeita$^{1,2}$, V\'i{}t Jakubsk\'y$^1$, Kevin Zelaya$^1$}

\affil[${1}$]{\footnotesize Nuclear Physics Institute, Czech Academy of Science, 250 68 \v{R}e\v{z}, Czech Republic}
\affil[${2}$]{\footnotesize
Physics Department, Cinvestav, P.O. Box. 14-740, 07000 Mexico City, Mexico }

\date{}

\begin{document}

\maketitle

\begin{abstract}
	
	We consider confinement of Dirac fermions in $AB$-stacked bilayer graphene by inhomogeneous on-site interactions, (pseudo-)magnetic field or inter-layer interaction. Working within the framework of four-band approximation, we focus on the systems where the stationary equation is reducible into two stationary equations with $2\times2$ Dirac-type Hamiltonians and auxiliary interactions. We show that it is possible to find localized states by solving an effective Schr\"odinger equation with energy-dependent potential. We consider several scenarios where bilayer graphene is subject to inhomogneous (pseudo-)magnetic field, on-site interactions or inter-layer coupling. In explicit examples, we provide analytical solutions for the states localized by local fluctuations or periodicity defects of the interactions.     
\end{abstract}


\section{Introduction}
In bilayer graphene, two flakes of graphene are close each other such that their electrons can mutually interact. The relative orientation of the two layers can vary. In case of Bernal (or $AB$-) stacking, the two layers are relatively shifted such that some bonds are parallel in the two lattices. Denoting the atoms in the two triangular sublattices of the $j$-th layer as $A_j$ and $B_j$, the atoms $A_2$ of the upper layer sit just above the $B_1$ atoms of the lower layer, whereas the atoms $A_1$ are below the centers of the hexagons of the upper lattice. The effective Hamiltonian (in the four-band approximation) for the low-energy particles can be written as \cite{McCann} 
\begin{equation}\label{Hblg0}
H_{blg}=\left(\begin{array}{cccc}
\epsilon_{A_1}&\pi^\dagger&0&v_3\pi\\
\pi&\epsilon_{B_1}&\gamma&0\\
0&\gamma&\epsilon_{A_2}&\pi^\dagger\\
v_3\pi^\dagger&0&\pi&\epsilon_{B_2}
\end{array}\right),\quad \pi=-i\partial_x-i\partial_y.
\end{equation}
The basis of the bispinors is $(A_1,B_1,A_2,B_2)$. The quantities $\epsilon_{A_j(B_j)}$, $j\in\{1,2\}$, correspond to the on-site energies that can originate from an external electric potential, spin-orbit interaction or an interaction with the substrate. The parameter $\gamma$ represents interaction of the electrons on the sites $A_2$ and $B_1$. The term proportional to $v_3$ is related to the interlayer trigonal warping \cite{McCann}, which is frequently set to zero in the literature. 
The four-band Hamiltonian (\ref{Hblg0}) was used in e.g. in the analysis of strains and their effect on electronic \cite{Mariani}, \cite{Verberck}, \cite{Cariglia} or topological properties of bilayer graphene \cite{Moulsdale}. It was used in the analysis of confinement of Dirac fermions in  quantum dots formed by doping \cite{MiltonPereira} or by local variation of the interlayer coupling related to local delamination of the bilayer graphene (graphene blisters) studied recently in \cite{Abdullah}, \cite{Solomon}. It serves well for description of other bilayer Dirac materials, see e.g. bilayer silicene  \cite{Ezawa}, \cite{Wu}, \cite{Ho}. Qualitatively the same operator with $v_3=0$ appears in description of spin-orbit interaction in graphene, see e.g. \cite{Rataj}, \cite{Esmaeilzadeh}.

Electrons on the binding sites $A_2$ and $B_1$ form dimers. For $E<<\gamma$, the dynamics on the non-dimer sites  $A_1$ and $B_2$ get dominant, and the effective Hamiltonian can be derived from (\ref{Hblg0}), see \cite{McCann}, \cite{Falco}. The latter case is known as the two-band approximation, in which the energy operator reads as $H=\left({}^0_{(\pi^\dagger)^2}\,\,{}^{\pi^2}_0\right)$. This framework proved to be useful in the analysis of various situations where electric or magnetic fields are inhomogeneous, or the bilayer is subjected to inhomogeneous deformations \cite{Falco}, \cite{Novoselov}, \cite{Katsnelson}. 
Let us also mention that the exactly solvable two-band Hamiltonians with inhomogeneous magnetic field were constructed \cite{Fernandez1}. Coherent states for the system described by this operator were found in \cite{Fernandez2}. In most cases, the studied systems possessed translational or rotational symmetry so that effectively one-dimensional systems were analyzed. 

In the current article, we are interested in systems where, besides the on-site interactions and (pseudo-)magnetic field, the inter-layer coupling $\gamma$ can also be inhomogeneous.  We are particularly interested in situations where fluctuations of the involved interactions can confine the Dirac fermions. In this quest, we prefer working within the framework of the four-band approximation. Therefore, we will consider the settings described by the following energy operator
\begin{equation}
H=\left(\begin{array}{cccc}
\epsilon_1(x)&-i\partial_x+A(x)&0&iv_3\partial_x\\
-i\partial_x+A^*(x)&\epsilon_2(x)&\gamma(x)&0\\
0&\gamma(x)&\epsilon_2(x)&-i\partial_x+A^*(x)\\
iv_3\partial_x&0&-i\partial_x+A(x)&\epsilon_1(x)
\end{array}\right).\label{redHblg}
\end{equation}
When compared to (\ref{Hblg0}), this operator can be matched with the Hamiltonian of bilayer graphene for inhomogeneous $\epsilon_{A_1}=\epsilon_{B_2}=\epsilon_1(x)$, $\epsilon_{B_1}=\epsilon_{A_2}=\epsilon_2(x)$ and with the longitudinal momentum $k_y=0$. Therefore, (\ref{redHblg}) can describe dynamics of the fermions that bounce on the potential in normal direction.  The operator (\ref{redHblg}) also contains an additional potential term $A_x\sigma_0\otimes\sigma_1-A_y\sigma_3\otimes\sigma_2$, $A=A_x+iA_y$. 

It was recently discussed in~\cite{Celeita} that the Hamiltonian (\ref{redHblg}) belongs to the class of reducible operators where the solution of the associated stationary equation can be found via two, lower-dimensional, dynamical equations with auxiliary interactions.  Indeed, let us make an ansatz for the wave functions
\begin{equation}\label{ansatz}
\mathbf{\Psi}=(\xi_1,\xi_2, \xi_2,\xi_1)^T,\quad \mathbf{\Xi}=(\chi_1,\chi_2,-\chi_2,-\chi_1)^T.
\end{equation}
Then, the bispinors $\mathbf{\Psi}$ and $\mathbf{\Xi}$ satisfy\footnote{Notice that we denote $a^*$ as complex conjugate of $a$ in the article.}
\begin{equation}
H\,\mathbf{\Psi}=E\,\mathbf{\Psi},\quad H\,\mathbf{\Xi}=\overline{E}\,\mathbf{\Xi}\label{stac},
\end{equation}
provided that the functions $\mathbf{\xi}=(\xi_1, \xi_2)^T$ and $\mathbf{\chi}=(\chi_1, \chi_2)^T$ are solutions of the following two equations, 
\begin{eqnarray}\label{reducedequationD}
&&h_1\,\xi=\left[-i\left(\begin{array}{cc}v_3&1\\1&0\end{array}\right)\partial_x+\left(\begin{array}{cc}\epsilon_1&A\\A^*& \epsilon_2+\gamma\end{array}\right)\right]\xi=E\,\xi,\label{eq1}\\
&& h_2\,\chi=\left[-i\left(\begin{array}{cc}-v_3&1\\1&0\end{array}\right)\partial_x+\left(\begin{array}{cc}\epsilon_1&A\\A^*& \epsilon_2-\gamma\end{array}\right)\right]\chi=\overline{E}\,\chi,\label{eq2}
\end{eqnarray}  
with $E$ and $\overline{E}$, in general, different from each other.

These equations resemble the $1+1$ dimensional Dirac equation up to the matrix coefficient in the kinetic term that contains the nonvanishing constant $v_3$. Let us investigate how the equations (\ref{eq1}) and (\ref{eq2}) can be solved and the physical scenarios that can be studied in this way.

\section{Bilayer graphene via Schr\"odinger equation with energy-dependent potential\label{sectiontwo}}
For the sake of generality, let us consider the matrix Hamiltonian
\begin{equation}
h=-i
\begin{pmatrix}
v_{3} & 1 \\
1 & 0
\end{pmatrix}
\partial_{x}+
\begin{pmatrix}
V_{1}(x) & A(x) \\
A^{*}(x) & V_{2}(x)
\end{pmatrix}
\, , \quad V_{1}(x),V_{2}(x):\mathbb{R}\rightarrow\mathbb{R} \, , \quad A(x):\mathbb{R}\rightarrow\mathbb{C} \, .\label{eq0}
\end{equation}
The stationary equation 
\begin{equation}\label{stacsmall}
h
\begin{pmatrix}
\psi_{1} \\ \psi_{2}
\end{pmatrix}
=E
\begin{pmatrix}
\psi_{1} \\ \psi_{2}
\end{pmatrix} \, ,
\end{equation}
gives rise to the set of  coupled equations for the spinor components $\psi_1$ and $\psi_2$,
\begin{align}\label{coupled1}
&-iv_{3}\psi_{1}'-i\psi_{2}'+A(x)\psi_{2}+(V_{1}(x)-E)\psi_{1}=0 \, , \\
&-i\psi_{1}'+A^{*}(x)\psi_{1}+(V_{2}(x)-E)\psi_{2}=0 \, ,\label{coupled2}
\end{align}
with $f'\equiv\partial f/\partial_{x}$.
The equations (\ref{coupled1})-(\ref{coupled2}) can be decoupled by fixing
\begin{equation}
\psi_{2}=\frac{-i\psi_{1}'+A^{*}(x)\psi_{1}}{E-V_{2}(x)} \, ,\label{psi2G}
\end{equation}
so that (\ref{coupled1}) turns into a second-order differential equation for $\psi_1$. In order to bring it into the Sturm-Liouville form, we make an additional energy-dependent transformation 
\begin{equation}
\psi_1(x)=e^{-i\phi(x)}\widetilde{\psi}_1(x) \, ,
\label{pahse1}
\end{equation}
where
\begin{equation}
\phi(x)=\frac{i}{2}\log (E-V_2(x))+\frac{1}{2}\int^{x}dx'\left(v_3(E-V_2(x))+2\operatorname{Re}A(x')\right) \, .
\end{equation}
Notice that the first term of $\phi(x)$ is not a pure phase so that the normalization of $\psi_1$ and $\widetilde{\psi}_{1}$ can differ. This should be kept in mind when imposing boundary conditions on the functions $\widetilde{\psi}_1$ in general. The equation for $\widetilde{\psi}_1$ reads as
\begin{align}
-\widetilde{\psi}_{1}''&+\left(-\operatorname{Im}A'(x)+(\operatorname{Im}A(x))^{2}-v_{3}(E-V_{2}(x))\operatorname{Re}A(x)\right.\nonumber\\&-\frac{v_3^2}{4}(E-V_2(x))^2
-(E-V_1(x))(E-V_2(x))\nonumber\\
&\left.-\frac{\operatorname{Im}A\, V_2'(x)}{E-V_2(x)}+\frac{3(V_2'(x))^2}{4(E-V_2(x))^2}+\frac{V_2''(x)}{2(E-V_2(x))}\right)\widetilde{\psi}_1=0 \, . 
\label{psi1t0}
\end{align}
Therefore, $\widetilde{\psi}_1$ can be found as the zero-mode of the Schr\"odinger equation with a potential given in terms of $A(x)$, $V_1(x)$, and $V_2(x)$. Although it is a rather complicated task to find its solutions in general, one can identify (\ref{psi1t0}) with a Schr\"odinger equation that possesses a solvable potential $V_0$, $-\widetilde{\psi}_1''+V_0\widetilde{\psi}_1=0$. However, by doing so, one has to solve a nonlinear differential equation for either $\operatorname{Im}A$ or $V_2$. To overcome this issue, one can fix either $\operatorname{Re}A(x)$ or $V_1(x)$ so that the potential in (\ref{psi1t0}) coincides with $V_0$. In this form, one of the two quantities could be fixed as
\begin{equation}
V_1=V_1(V_2,A,V_0,E)\quad \mbox{or}\quad \operatorname{Re}A=\operatorname{Re}A(V_2,V_1,\operatorname{Im}A,V_0,E).
\end{equation}
This way, we can acquire a solution $\widetilde{\psi}_1$ with a single energy level $E$. Indeed, changing $E$ would alter the interaction $V_1(x)$ so that we would deal with a different physical setting. We will discuss this situation in the section \ref{four}.

For constant $V_2$, Eq.~(\ref{psi1t0}) simplifies considerably as the third line in (\ref{psi1t0}) cancels out, leading to
\begin{equation}
-\widetilde{\psi}_{1}''+{V}_{E}(x) \widetilde{\psi}_{1} = \lambda_{E} \widetilde{\psi}_{1} \, ,
\label{psi1t}
\end{equation}
where
\begin{align}
{V}_{E}(x)&=-\operatorname{Im}A'(x)+(\operatorname{Im}A(x))^{2}-v_{3}(E-V_{2})\operatorname{Re}A(x) -(E-V_{2})(E-V_{1}(x))\, , 
\label{schro1}\\ 
\lambda_{E}&=\frac{v_{3}^{2}}{4}(E-V_{2})^{2}\, .
\label{lambda1}
\end{align}
The equation~\eqref{psi1t} corresponds to a stationary Schr\"odinger equation with an \textit{energy-dependent potential} term ${V}_{E}(x)$. The energy-dependent part in (\ref{schro1}) cancels out effectively when $V_1$ is constant and either $v_{3}$ vanishes or Re$A$ is a constant. In the later case, the term $v_{3}(E-V_2)A_{0}+(E-V_1)(E-V_2)$ adds a shift to the eigenvalue $\lambda_{E}$. 

Notice that when $E=V_2$, the equations~(\ref{coupled1})-(\ref{coupled2}) can be solved directly with
\begin{equation}\label{V2=E}
\psi_1=e^{-i\int A^*(x)dx},\quad \psi_2=-\frac{i\int e^{i\int( A(x)-A^*(x))dx}(V_1-V_2(x)-v_3A^*(x))dx}{e^{i\int A(x)dx}}.
\end{equation}

We can recover the $2\times 2$ bilayer Hamiltonians $h_{1}$ and $h_{2}$ in (\ref{eq1}) and (\ref{eq2}), respectively, through the following identification,
\begin{equation}\label{identification}
h_{1}\equiv h\vert_{\substack{V_{1}=\epsilon_1 \\ V_{2}=\epsilon_2+\gamma}} \, , \quad h_{2}\equiv h\vert_{\substack{V_{1}=\epsilon_1 \\ V_{2}=\epsilon_2-\gamma \\ v_{3}\rightarrow -v_{3}}} \, .
\end{equation}
Therefore, if we have
\begin{equation}\label{hstac}
h\left({}^{\psi_{1}}_{\psi_{2}}\right)=E\left({}^{\psi_{1}}_{\psi_{2}}\right),
\end{equation}
then there also holds
\begin{align}
h_{1}\xi=\varepsilon\,\xi&,\quad\xi=\left({}^{\xi_{1}}_{\xi_{2}}\right)\equiv\left({}^{\psi_{1}}_{\psi_{2}}\right)\vert_{\substack{V_{1}=\epsilon_{1}\\V_{2}=\epsilon_{2}+\gamma}} \, ,\quad \varepsilon\equiv E\vert_{\substack{V_{1}=\epsilon_{1}\\V_{2}=\epsilon_{2}+\gamma}}, \label{hom-spin1} \\  h_{2}\chi=\widetilde{\varepsilon}\,\chi&,\quad\chi=\left({}^{\chi_{1}}_{\chi_{2}}\right)\equiv\left({}^{\psi_{1}}_{\psi_{2}}\right)\vert_{\substack{V_{1}=\epsilon_{1}\\V_{2}=\epsilon_{2}-\gamma\\v_{3}\rightarrow-v_{3}}} \, ,\quad 
\widetilde{\varepsilon}\equiv E\vert_{\substack{V_{1}=\epsilon_{1}\\V_{2}=\epsilon_{2}-\gamma\\v_{3}\rightarrow-v3}} \, .
\label{hom-spin2}
\end{align}
The corresponding bispinor solutions of (\ref{stac}) are
\begin{align}
\mathbf{\Psi}&=\frac{1}{\sqrt{2}}\left(\xi,\sigma_1\xi\right)=\frac{1}{\sqrt{2}}\left(\psi_1,\frac{-i\psi_{1}'+A^{*}(x)\psi_{1}}{E-V_{2}} ,\frac{-i\psi_{1}'+A^{*}(x)\psi_{1}}{E-V_{2}} ,\psi_1 \right)^{T}\vert_{\substack{V_{1}=\epsilon_1 \\ V_{2}=\epsilon_2+\gamma}},\nonumber\\
\mathbf{\Xi}&=\frac{1}{\sqrt{2}}\left(\chi,-\sigma_1\chi\right)=\frac{1}{\sqrt{2}}\left(\psi_1,\frac{-i\psi_{1}'+A^{*}(x)\psi_{1}}{E-V_{2}} ,-\frac{-i\psi_{1}'+A^{*}(x)\psi_{1}}{E-V_{2}} ,-\psi_1 \right)^{T} \vert_{\substack{V_{1}=\epsilon_1 \\ V_{2}=\epsilon_2-\gamma \\ v_{3}\rightarrow -v_{3}}}
\, . \label{PsiXi}\quad
\end{align}
The bispinors are normalized provided that $\psi$ and $\chi$ are normalized. Let us notice that if we can find a fundamental systems of solutions of (\ref{eq1}) and (\ref{eq2}) for $E=\varepsilon$, then we can find the fundamental system for (\ref{stac}) for $E=\widetilde{\varepsilon}$. 
 
When $\epsilon_2$ and $\gamma$ are constants, both $h_1$ and $h_2$ correspond to $h$ with constant $V_1$ and $V_2$ (recall that the explicit form of $V_2$ differs in $h_1$ and $h_2$). In this case, the potential ${V}_E$ and ${\lambda}_E$ in (\ref{psi1t}) acquire this form
\begin{align}
{V}_{E}(x)&=-\operatorname{Im}A'(x)+(\operatorname{Im}A(x))^{2}-v_{3}(E-V_{2})\operatorname{Re}A(x)\, , 
\label{schro1b}\\ 
\lambda_{E}&=\frac{v_{3}^{2}}{4}(E-V_{2})^{2}+(E-V_{2})(E-V_{1})\,
\label{lambda1b}.
\end{align}
As $V_2$ is constant, the transformation (\ref{pahse1}) does not alter integrability, such that when $\widetilde{\psi}_1$ is square-integrable, so is $\psi_1=e^{-i\phi(x)}\widetilde{\psi}_1$. Therefore, we get solutions of (\ref{eq1}) and (\ref{eq2}) as in (\ref{hom-spin1}) and (\ref{hom-spin2}), respectively. The two solutions (\ref{ansatz}) for the stationary equation (\ref{stac}) then read as in (\ref{PsiXi}).

Now, if we allow $\epsilon_1$, $\epsilon_2$ and $\gamma$ to be inhomogeneous, we can still keep $V_2$ constant in either $h_1$ or $h_2$, provided either $\epsilon_2+\gamma$ or $\epsilon_2-\gamma$ is constant, respectively. For convenience, let us fix that there holds\footnote{In the case $V_2=\epsilon_2(x)-\gamma(x),$ $V_2\in\mathbb{R}$, all the analysis below is applicable with minor changes.}
\begin{equation}\label{gamma}
V_2=\epsilon_2(x)+\gamma(x),\quad V_2\in\mathbb{R}.
\end{equation}
In this case, we get constant $V_2=\epsilon_2(x)+\gamma(x)$ in $h_1$  whereas $h_2$ corresponds to $h$ with inhomogeneous $V_2(x)=\epsilon_2(x)-\gamma(x)$, see (\ref{identification}). Notice that $V_1$ can be inhomogeneous now. The equation (\ref{eq1}) reduces into (\ref{psi1t}) whereas (\ref{eq2}) leads to (\ref{psi1t0}). It is feasible to solve analytically only one of the two equations by fixing $\epsilon_1(x)$ appropriately. In either case, the analytical solutions of (\ref{stac}) are either $\mathbf{\Psi}$ or $\mathbf{\Xi}$ in (\ref{PsiXi}) with $V_1\rightarrow \epsilon_1(x)$.
It is worth noticing that the solutions of (\ref{psi1t}) depend just on $\epsilon_2(x)+\gamma(x)$, they do not "feel" the explicit form of the inter-layer interaction $\gamma(x)$ and of the on-site coupling $\epsilon_2(x)$. These interactions can be changed without altering $\widetilde{\psi}_1$ provided the change complies with (\ref{gamma}).
As we can find only part of the solutions of (\ref{stac}) analytically, we consider the models with inhomogeneous $\epsilon_2$ and $\gamma$ that satisfy (\ref{gamma}) as quasi-exactly solvable. 

Solution of (\ref{psi1t}) with the energy-dependent potential (\ref{schro1}) is nontrivial. Identification of a complete set of solutions to a energy-dependent Schr\"odinger equation is, if possible, a challenging task in most cases~\cite{Formanek}, \cite{Schulze}. We will focus on the cases where confinement of Dirac fermions is caused either by $A(x)$, $\epsilon_1(x)$, or $\gamma(x)$. The three situations will be discussed in the following three sections separately. Our strategy will be to identify the potential (\ref{schro1}) with the potential of a known solvable system.  This way, we will be able to identify a set of square-integrable solutions and the corresponding set of eigenvalues $\lambda_{E}$. 

\section{Confinement by the vector potential\label{three}}
\label{conf-vec-pot}
The potential term $A_x\sigma_0\otimes\sigma_1-A_y\sigma_3\otimes\sigma_2$ in (\ref{Hblg0}) resembles a magnetic vector potential. However, the corresponding magnetic field would have a different sign on the two layers, which might be physically unfeasible. Nevertheless, it is known that deformations of graphene layers are manifested in the form of the effective (pseudo-)magnetic vector potential \cite{Mariani}, \cite{Naumis}, \cite{Moldovan}. Therefore, we can see $A$ as a combination of the magnetic and pseudo-magnetic field that can acquire different values on the two layers, see e.g. \cite{Crosse}.

In this section, we fix $\epsilon_1$, $\epsilon_2$ as well as $\gamma$ in (\ref{redHblg}) to be constant,
\begin{equation}
\epsilon_1,\ \epsilon_2,\ \gamma\in\mathbb{R}.
\end{equation}
It allows us to convert the task of solving both (\ref{eq1}) and (\ref{eq2}) into solution of the Schr\"odinger equation with energy-dependent potential (\ref{psi1t}). We will identify the latter equation with a stationary equation of known exactly solvable model. This way, we shall find the explicit solutions $\widetilde{\psi}_1$ and energies $E$ of (\ref{psi1t}). As a consequence, the corresponding solutions of (\ref{eq1}) and (\ref{eq2}) are found through the relationships~\eqref{hom-spin1} and~\eqref{hom-spin2}, respectively. We will discuss two settings, the harmonic oscillator and Rosen-Morse system.

\subsection{Harmonic oscillator case}
Let us fix the vector potential $A(x)$ as a complex-valued function linear in $x$ in both its real and imaginary parts, so that
\begin{equation}
A(x)=A_{0}+m(w_{r}+i\,w_{i})x \, , \quad m,w_{r},w_{i} \in \mathbb{R} \, .
\label{AOsc}
\end{equation}
We can interpret the vector potential term as the consequence of external homogeneous magnetic field and a mechanical strain that gives rise to homogeneous pseudo-magnetic field. Homogeneous pseudo-magnetic field in bilayer graphene was discussed in \cite{Moldovan}, whereas Dirac fermions in bilayer graphene in presence of homogeneous magnetic field were discussed in \cite{Cariglia}, \cite{Falco}.

The effective potential  $V_{E}$ in~\eqref{psi1t} takes the form
\begin{equation}\label{Vho}
{V}_E(x)=m^2w_{i}^{2}x^{2}-mw_{r}v_{3}(E-V_{2})x-mw_{i}-v_{3}A_{0}(E-V_{2}) \, .
\end{equation}
Clearly, the latter implies that we must solve the eigenvalue equation of the well-known stationary  oscillator. Its general solution can be found once we cast the eigenvalue equation into the confluent hypergeometric equation. It reads explicitly as (for details, see~\cite{Nik88})
\begin{multline}
\widetilde{\psi}_{1}=e^{-\frac{mw_{i}}{2} \left( x -\frac{\Omega_{E}}{w_{i}} \right)^{2}} \left( \ell_{1} \, {}_{1}F_{1}\left[ \frac{1}{4}-\frac{\lambda_{E}}{4mw_{i}} , \frac{1}{2} ; m w_{i} \left( x-\frac{\Omega_{E}}{w_{i}} \right)^{2} \right] \right. \\
+ \left. \ell_{2} \left( x-\frac{\Omega_{E}}{w_{i}}\right) \, {}_{1}F_{1}\left[ \frac{3}{4}-\frac{\lambda_{E}}{4mw_{i}} , \frac{3}{2} ; m w_{i} \left( x-\frac{\Omega_{E}}{w_{i}} \right)^{2} \right] \right) \, .
\label{oscgen1}
\end{multline}
Here, $\ell_1$ and $\ell_2$ are constant coefficients, ${}_{1}F_{1}[a,b;z]$ stands for the \textit{Kummer or confluent hypergeometric} function~\cite{Olv10}, and 
\begin{equation}
\Omega_{E}:=\frac{w_{r}v_{3}}{2mw_{i}}(E-V_{2}) \, , \quad  \lambda_{E}=\frac{v_{3}^{2}}{4}\left(1+\frac{w_{r}^{2}}{w_{i}^{2}} \right)^{2}(E-V_{2})^{2}+(E-V_{2})(E-V_{1}+v_{3}A_{0})+mw_{i} \, .
\label{losc1}
\end{equation}
It is worth noticing that the energy-dependent terms in the potential (\ref{Vho}) cause just shifted decentering $\Omega_E$ of the harmonic oscillator and shift the energies $\lambda_E$.

Now, from the asymptotic behavior of the hypergeometric functions, it is straightforward to determine the physical values $E$ for which the function $\widetilde{\psi}_{1}$ becomes square integrable. This is achieved by imposing a polynomial behavior on the confluent hypergeometric function so that the Gaussian term~(\ref{oscgen1}) vanishes faster than the polynomial at $x\rightarrow\pm\infty$. This leads to an exponentially vanishing function for large $|x|$. Such a polynomial behavior is achieved if $a=-n$ in ${}_{1}F_{1}[a,b;z]$, with $n$ being a non-negative integer or zero. To simplify the discussion we consider two cases. First, the conditions $\ell_{2}=0$ and $\left(\frac{1}{4}-\frac{\lambda_{E}}{4mw_{i}}\right)=-n$ lead to $\lambda_{E}=2mw_{i}(2n+1/2)$, from which we obtain $\widetilde{\psi}_{1}$ in terms of even Hermite polynomials $H_{2n}(z)$. Second, the conditions $\ell_{1}=0$ and $\left(\frac{3}{4}-\frac{\lambda_{E}}{4mw_{i}}\right)=-n$ allow us to obtain $\lambda_{E}=2mw_{i}((2n+1)+1/2)$, and solutions for $\widetilde{\psi}_{1}$ in terms of odd Hermite polynomials $H_{2n+1}(z)$. 

We can thus unify both the even and odd solutions as
\begin{equation}
\widetilde{\psi}_{1;n}\equiv\widetilde{\psi}_{1}=e^{-\frac{mw_{i}}{2}\left(x-\frac{\Omega_{E}}{w_{i}}\right)^{2}} \, H_{n}\left[ \sqrt{mw_{i}} \left( x - \frac{\Omega_{E}}{w_{i}}\right) \right] \, , \quad \lambda_{E}=2mw_{i}\left(n+\frac{1}{2}\right) \, .
\label{losc2}
\end{equation}
The physical energies $E_{n}$ are determined after comparing $\lambda_E$ in~\eqref{losc1} with~\eqref{losc2}. We get
\begin{multline}
E^{(\pm)}_{n}=\frac{2(V_{1}+V_{2})-2v_{3}A_{0}+V_{2}v_{3}^{2}\left( 1+\frac{w_{r}^2}{w_{i}^2}\right)\pm 2\sqrt{(V_{1}-V_{2}-v_{3}A_{0})^{2}+2nmw_{i}\left( 4+v_{3}^2\left(1+\frac{w_{r}^{2}}{w_{i}^{2}} \right) \right)}}{4+v_{3}^2\left(1+\frac{w_{r}^{2}}{w_{i}^{2}}\right)}.
\end{multline}

Here, it is worth remarking that the case $n=0$ should be addressed with caution, as the exact value of $E_{0}^{(\pm)}$ depends on the sign of $V_{1}-V_{2}-v_{3}A_{0}$. That is,
\begin{equation}
\begin{aligned}
&E_{0}^{(+)}=
\begin{cases}
V_{2}+4\frac{V_{1}-V_{2}-v_{3}A_{0}}{4+v_{3}^{2}\left(1+\frac{w_{r}^{2}}{w_{i}^{2}} \right)}, \quad &  \, V_{1}-V_{2}-v_{3}A_{0}>0 \\
V_{2},  \quad &  \, V_{1}-V_{2}-v_{3}A_{0}<0
\end{cases}
\, , \\
&E_{0}^{(-)}=
\begin{cases}
V_{2},  \quad &  \, V_{1}-V_{2}-v_{3}A_{0}>0 \\
V_{2}+4\frac{V_{1}-V_{2}-v_{3}A_{0}}{4+v_{3}^{2}\left(1+\frac{w_{r}^{2}}{w_{i}^{2}} \right)}, \quad &  \, V_{1}-V_{2}-v_{3}A_{0}<0
\end{cases} 
\, ,
\end{aligned}
\label{E0osc}
\end{equation}
from which we see that $E=V_{2}$ for either $E_{0}^{(+)}$ or $E_{0}^{(-)}$. We can find the spinor corresponding to this energy level from~\eqref{V2=E}. It reads as 
\begin{align}
& \left.\psi_{0}^{(\pm)}\right\vert_{E=V_{2}}=\mathcal{N}_{0}^{(\pm)}e^{-i\left(A_{0}x+\frac{m w_{r}}{2}x^2 \right)}e^{-\frac{m w_{i}}{2}x^{2}}
\begin{pmatrix}
1 \\
\sqrt{\frac{\pi}{mw_{i}}}\frac{V_{1}-V_{2}-v_{3}A_{0}}{2}e^{mw_{i}x^{2}}\operatorname{Erf}(\sqrt{m w_{i}}\,x)+\frac{v_{3}}{2}\left(\frac{w_{r}}{w_{i}}-i\right)
\end{pmatrix}
\,.
\end{align}
It is not square-integrable as its the second component diverges for $x\rightarrow\pm\infty$. The latter means that $E=V_{2}$ is not a physical energy, and we thus identify the point spectrum of $h$ as
\begin{equation}
\operatorname{Sp}(h)=
\begin{cases}
\{E_{n}^{(+)}\}_{n=0}^{\infty}\cup\{ E_{n+1}^{(-)} \}_{n=0}^{\infty} \, , \quad  & V_{1}-V_{2}-v_{3}A_{0}>0 \\
\{E_{n+1}^{(+)}\}_{n=0}^{\infty}\cup\{ E_{n}^{(-)} \}_{n=0}^{\infty} \, , \quad  & V_{1}-V_{2}-v_{3}A_{0}<0
\end{cases}
\, .
\label{Avec-osc-spec}
\end{equation}

Now, we identify the corresponding spinor to each element in Sp$(h)$. From the general solution~\eqref{oscgen1}, and after some calculations, we get
\begin{multline}
{\psi}^{(\pm)}_{n}=
\begin{pmatrix}
\psi_{1;n}^{(\pm)} \\
\psi_{2;n}^{(\pm)}
\end{pmatrix}
=\mathcal{N}_{n}^{(\pm)}e^{-i\left(\frac{v_3}{2}(E_{n}^{(\pm)}-V_{2})+A_{0}\right)x}e^{-i\frac{mw_{r}}{2}x^{2}}e^{-\frac{(z_{n}^{(\pm)}(x))^{2}}{2}} \times \\
\begin{pmatrix}
H_{n}(z^{(\pm)}_{n}(x)) \\
-\frac{v_{3}}{2}\left(1+i\frac{w_{r}}{w_{i}}\right)H_{n}(z_{n}(x))-2in\frac{\sqrt{mw_{i}}}{E_{n}^{(\pm)}-V_{2}}H_{n-1}(z^{(\pm)}_{n}(x))
\end{pmatrix} 
\, ,
\label{Avec-osc-sols}
\end{multline}
for $n=0,1,\ldots$, and $E_{0}^{(\pm)}\not=V_{2}$, where we have introduced the reparametrized coordinate and decentering shift
\begin{equation}
z_{n}^{(\pm)}(x)=\sqrt{mw_{i}} \left( x - \frac{\Omega^{(\pm)}_{n}}{w_{i}}\right) \, , \quad \Omega^{(\pm)}_{n}:=\frac{w_{r}v_{3}}{2mw_{i}}(E_{n}^{(\pm)}-V_{2}) \, ,
\end{equation}
respectively.

Although the Hermite polynomials in~\eqref{Avec-osc-sols} depend explicitly on the energy, it is still possible to compute the normalization factor for each spinor. This is done by exploiting the well-known properties of the Hermite polynomials, leading to, up to a global complex-phase,
\begin{equation}
\mathcal{N}_{n}^{(\pm)}= \sqrt{\frac{1}{2^{n}n!}\sqrt{\frac{m w_{i}}{\pi}}}\left(1+\frac{v_{3}^{2}}{4}\left(1+\frac{w_{r}^{2}}{w_{i}^{2}} \right)^{2}+\frac{2n m w_{i}}{(E_{n}^{(\pm)}-V_{2})^{2}} \right)^{-\frac{1}{2}} \, , \quad \, n=0,1\ldots \, ,
\label{Avec-osc-norm}
\end{equation}
which holds true only for the elements in~\eqref{Avec-osc-spec}. 

Heretofore, we have determined the eigenvalue problem related to $h$, and now the corresponding information for the reduced Hamiltonians $h_{1}$ and $h_{2}$ may be extracted directly from the relationships given in~\eqref{hom-spin1}--\eqref{hom-spin2}. For clarity, we use the notation 
\begin{equation}
\varepsilon_{n}^{(\pm)}=\left.E_{n}^{(\pm)}\right\vert_{\substack{V_{1}=\epsilon_{1}\\V_{2}=\epsilon_{2}+\gamma}} \, , \quad 
\widetilde{\varepsilon}_{n}^{(\pm)}=\left.E_{n}^{(\pm)}\right\vert_{\substack{V_{1}=\epsilon_{1}\\V_{2}=\epsilon_{2}-\gamma \\v_{3}\rightarrow -v_{3}}} \, ,
\label{Avec-osc-en}
\end{equation}
to denote the physical energies of $h_{1}$ and $h_{2}$, respectively. The corresponding bispinors $\mathbf{\Psi}_{n}$ and $\mathbf{\Xi}_{n}$ of (\ref{redHblg}) are similarly extracted via (\ref{PsiXi}). 

\begin{figure}
	\centering
	\subfloat[][Probability density for $\mathbf{\Psi}_{n}^{(\pm)}$]{\includegraphics[width=0.4\textwidth]{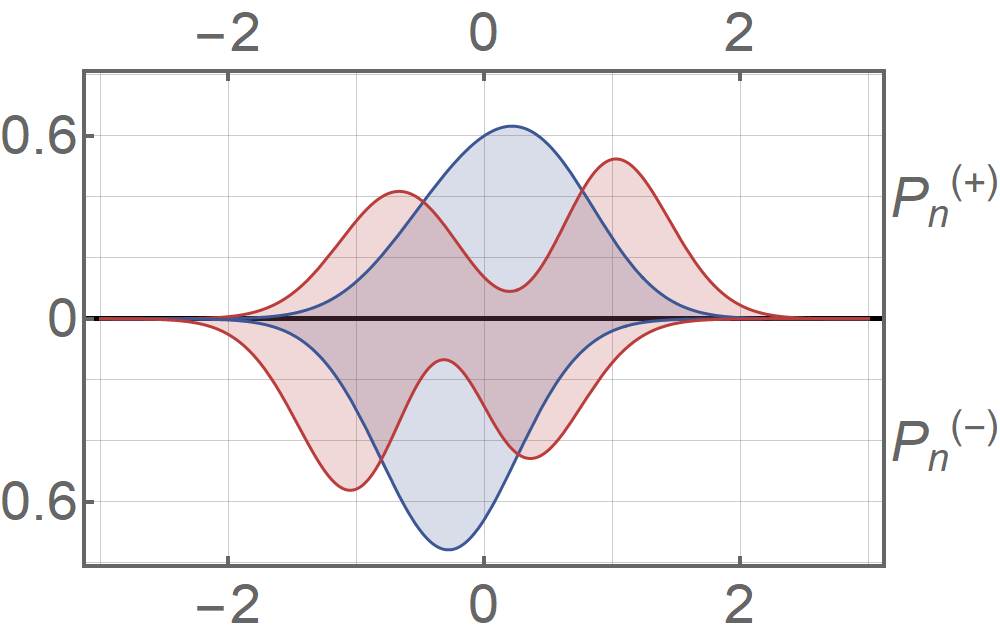}
		\label{Avec-osc-pd1}}
	\hspace{5mm}
	\subfloat[][Probability density for $\mathbf{\Xi}_{n}^{(\pm)}$]{\includegraphics[width=0.4\textwidth]{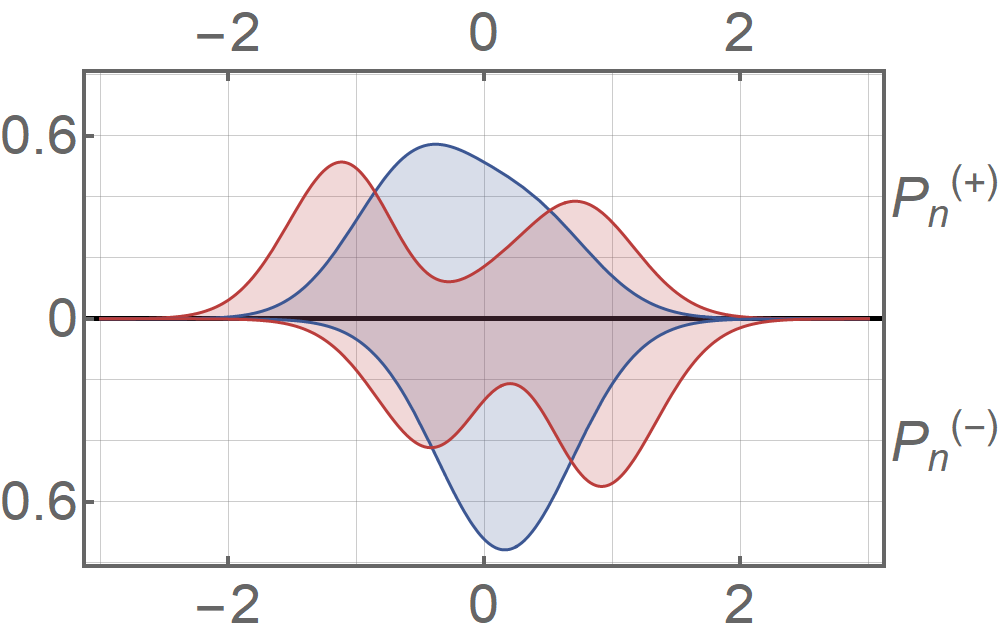}
		\label{Avec-osc-pd2}}
	\\
	\subfloat[][]{\includegraphics[width=0.4\textwidth]{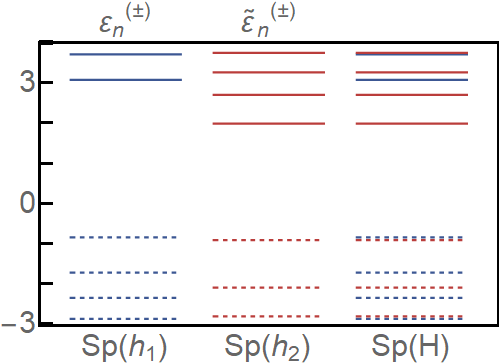}}
	\caption{(a) Probability densities $\mathcal{P}_{n}^{(\pm)}$ related to the bispinors $\mathbf{\Psi}_{n}^{(\pm)}$ for $n=1$ (blue) and $n=2$ (red). (b) Probability densities $\mathcal{P}_{n}^{(\pm)}$ related to the bispinors $\mathbf{\Xi}_{n}^{(\pm)}$ for $n=0$ (blue) and $n=1$ (red). (c) Energy levels $\varepsilon^{(+)}_{n}$ (blue-solid), $\varepsilon^{(-)}_{n}$ (blue-dashed), $\widetilde{\varepsilon}^{(+)}_{n}$ (red-solid), and $\widetilde{\varepsilon}^{(-)}_{n}$ (red-dashed). In all the cases, the parameters have been fixed as $m=1$, $A_{0}=0$, $w_{r}=2$, $w_{i}=1.8$, $\epsilon_1=-1$, $\epsilon_2=1.5$, $v_{3}=0.3$, and $\gamma=0.7$.}
	\label{SpOsc2}
\end{figure}

Particularly, we depict the energy levels structure of the bilayer Hamitlonian $H$ in Fig.~\ref{SpOsc2}, for $m=1$, $A_{0}=0$, $\epsilon_{1}=-1$, $w_{r}=2$, $w_{i}=1.8$, $\epsilon_{2}=1.5$, $v_{3}=0.3$, and $\gamma=0.7$. In this configuration, $V_{1}-V_{2}-A_{0}v_{3}<0$, and so the energies $\varepsilon_{0}^{(+)}$ and $\widetilde{\varepsilon}_{0}^{(+)}$ are both removed from the spectrum. Similarly, the corresponding bispinors are discarded. The behavior of the probability distributions related to the bispinors $\mathbf{\Psi}_{n+1}^{(+)}$, $\mathbf{\Xi}_{n+1}^{(+)}$, $\mathbf{\Psi}_{n}^{(-)}$, and $\mathbf{\Xi}_{n}^{(-)}$ are depicted in Figs.~\ref{Avec-osc-pd1}-\ref{Avec-osc-pd2} for $n=0,1$.

\subsection{Morse-Rosen potential}
\label{A-Morse-Rosen}
Now, let us associate $A(x)$ with a smooth step-like profile, defined in terms of a purely imaginary function of the form,
\begin{equation}\label{RMA}
A(x)=iU_{0}\left( \frac{\ell(\ell-1)}{2(\kappa-1)}+(\kappa-1)\operatorname{tanh}(U_{0}x) \right) \, , \quad \ell,\kappa\geq 1 \, .
\end{equation}
In this case, the effective potential in~\eqref{psi1t} becomes
\begin{equation}
{V}_{E}(x)=-U_{0}^{2}\kappa(\kappa-1)[\operatorname{sech}(U_{0}x)]^{2}+U_{0}^{2}\ell(\ell-1)\operatorname{tanh}(U_{0}x) + U_{0}^{2}\left[\frac{\ell^{2}(\ell-1)^{2}}{4(1-\kappa)^{2}}+(1-\kappa)^{2}\right] \, ,
\label{pot-RM}
\end{equation}
which corresponds to the \textit{Morse-Rosen interaction}~\cite{Bar87} (also known as the \textit{hyperbolic Rosen-Morse} potential). Interestingly, for $\ell=1$, the potential~\eqref{pot-RM} reduces to the \textit{P\"oschl-Teller} interaction, a particular case to be discussed in detail in the upcoming sections. Let us notice that Dirac fermions in (single-layer) graphene in presence of (\ref{RMA}) were discussed in \cite{Milpas}.

For the rest of this section, we focus on the case $\ell\geq1$  and $\kappa\not=1$. We will consider bound states that comply with the boundary condition $\left.\widetilde{\psi}_{1}\right\vert_{x\rightarrow\pm \infty}\rightarrow 0$. Thus, taking the differential equation for $\widetilde{\psi}_{1}$ into the hypergeometric form~\cite{Nik88}, and after some calculations, we get the eigenstates
\begin{align}
& \widetilde{\psi}_{1;n}=\mathcal{C}_{n}^{(\pm)}\left(1-\operatorname{tanh}(U_{0}x) \right)^{\frac{\alpha_{n}}{2}} \left(1+\operatorname{tanh}(U_{0}x) \right)^{\frac{\beta_{n}}{2}} P_{n}^{(\alpha_{n},\beta_{n})}(\operatorname{tanh}(U_{0}x)) \, ,
\label{RMboundstates} \\
& \lambda_{E}=-U_{0}^{2}(\beta_{n}^{2}-\beta_{0}^{2})=-U_{0}^{2}\left(\frac{\ell^{2}(\ell-1)^{2}}{4(n+1-\kappa)^{2}}+(n+1-\kappa)^{2}-\frac{\ell^{2}(\ell-1)^{2}}{4(1-\kappa)^{2}}-(1-\kappa)^{2} \right) \, ,
\label{RMenergies}
\end{align}
where $n=0,1,\ldots,n_{max}$, and $P_{n}^{(\alpha,\beta)}(z)$ stands for the \textit{Jacobi polynomials}, with
\begin{equation}
\alpha_{n}:= (\kappa-n-1)-\frac{\ell(\ell-1)}{2(n+1-\kappa)} \, , \quad \beta_{n}:= (\kappa-n-1)+\frac{\ell(\ell-1)}{2(n+1-\kappa)} \, .
\end{equation}
From~\eqref{RMboundstates}, it follows that $\widetilde{\psi}_{1;n}$ is square-integrable only when both $\alpha_{n},\beta_{n}>0$. This leads us to a condition for the existence of at least one bound state and an upper bound $n_{max}$ given by
\begin{equation}
(\kappa-1)^{2}>\frac{\ell(\ell-1)}{2} \, , \quad n_{max}=\left\lfloor\kappa-1-\sqrt{\frac{\ell(\ell-1)}{2}} \right\rfloor \, ,
\label{cond2-RM}
\end{equation}
respectively.

The physical energies $E_{n}$ are then determined by comparing $\lambda_{E}$ in~\eqref{RMenergies} with~\eqref{lambda1b}, from which we obtain a polynomial equation of second-order for $E_{n}$. We thus get the energies
\begin{equation}
E_{n}^{(\pm)}=\frac{2(V_{1}+V_{2})+v_{3}^{2}V_{2}\pm 2 \sqrt{(V_{1}-V_{2})^{2}+U_{0}^{2}(4+v_{3}^{2})(\beta_{0}^{2}-\beta_{n}^{2})}}{4+v_{3}^{2}} \, .
\end{equation}
The second spinor component $\psi_{2;n}$ is determined from~\eqref{psi2G}, and the corresponding spinors take the form
\begin{multline}
{\psi}^{(\pm)}_{n}=
\begin{pmatrix}
\psi_{1;n}^{(\pm)} \\
\psi_{2;n}^{(\pm)}
\end{pmatrix}
=e^{-i\frac{v_{3}}{2}(E_{n}^{(\pm)}-V_{2})x}(1-z)^{\frac{\alpha_{n}}{2}}(1+z)^{\frac{\beta_{n}}{2}} \times \\
\begin{pmatrix}
P_{n}^{(\alpha_{n},\beta_{n})}(z) \\
-\left(\frac{v_{3}}{2}+\frac{iU_{0}}{(E_{n}^{(\pm)}-V_{2})}\left( \frac{\ell(\ell-1)}{2(\kappa-1)} + \frac{\ell(\ell-1)}{2(n+1-\kappa)}+nz\right)\right)P_{n}^{(\alpha_{n},\beta_{n})}(z)-i \frac{U_{0}(2\kappa-1-n)(1-z^{2})}{2(E_{n}^{(\pm)}-V_{2})}P_{n-1}^{(\alpha_{n}+1,\beta_{n}+1)}(z)
\end{pmatrix}
\, ,
\end{multline}
with $z\equiv z(x):=\operatorname{tanh}(U_{0}x)$. 

The exact value of $E_{0}^{(\pm)}$ depends on the sign of $V_{1}-V_{2}$. In analogy to the oscillator-like interaction of the previous section, we have $E_{0}^{(+)}=V_{2}$ for $V_{1}<V_{2}$, and $E_{0}^{(-)}=V_{2}$ for $V_{1}>V_{2}$. The corresponding spinors (\ref{V2=E}) are not square integrable so that these values do not belong to the point spectrum of $h$. 

We found the following set $\operatorname{Sp}(h)$ of discrete energies of $h$,
\begin{align}
\operatorname{Sp}(h)=
\begin{cases}
&\{E_{n}^{(+)}\}_{n=1}^{n_{max}}\cup\{E_{n}^{(-}\}_{n=0}^{n_{max}} \, , \quad V_{1}<V_{2} \\
&\{E_{n}^{(+)}\}_{n=0}^{n_{max}}\cup\{E_{n}^{(-}\}_{n=1}^{n_{max}} \, , \quad V_{1}>V_{2}
\end{cases}
\, .
\label{Avec-RM-spec}
\end{align}

Since $\alpha_{n}$ and $\beta_{n}$ do not depend on the energy upper-index $(\pm)$, the upper bound $n_{max}$ is the same for both energies $E_{n}^{(\pm)}$. For $n_{max}=0$, the set of discrete energies is just $\{E_{0}^{(-)}\}$ for $V_{1}<V_{2}$, and $\{E_{0}^{(+)}\}$ for $V_{1}>V_{2}$. Moreover, the point spectrum may be empty if the inequality in~\eqref{cond2-RM} is not fulfilled. 

Now, the discrete energy levels associated with $h_1$ and $h_2$ are obtained via (\ref{hom-spin1}) and (\ref{hom-spin2}), 

\begin{equation*}
\operatorname{Sp}(h_1)=\operatorname{Sp}(h)\vert_{\substack{V_{1}=\epsilon_1 \\ V_{2}=\epsilon_2+\gamma}},\quad \operatorname{Sp}(h_2)=\operatorname{Sp}(h)\vert_{\substack{V_{1}=\epsilon_{1}\\V_{2}=\epsilon_{2}-\gamma\\v_{3}\rightarrow-v3}}.
\end{equation*}

\begin{figure}
	\centering
	\subfloat[][Probability density for $\mathbf{\Psi}_{n}^{(\pm)}$]{\includegraphics[width=0.3\textwidth]{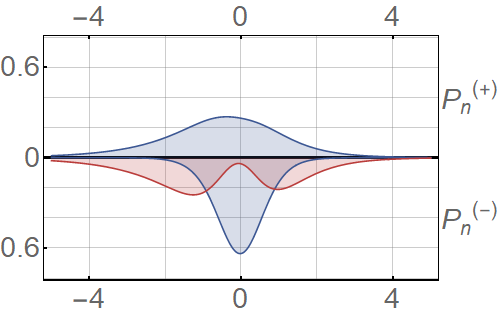}
		\label{Avec-RM-pd1}}
	\hspace{5mm}
	\subfloat[][Probability density for $\mathbf{\Xi}_{n}^{(\pm)}$]{\includegraphics[width=0.3\textwidth]{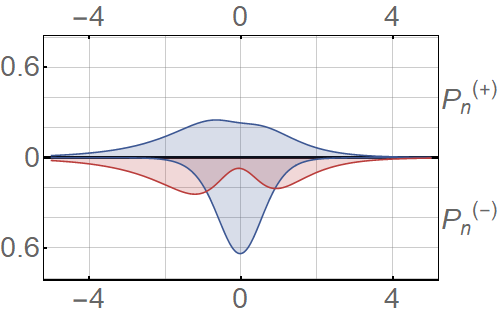}
		\label{Avec-RM-pd2}}
	\hspace{5mm}
	\subfloat[][]{\includegraphics[width=0.3\textwidth]{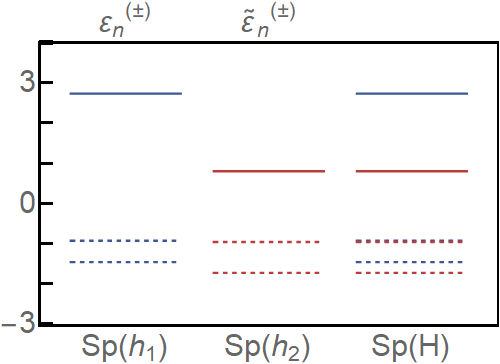}
		\label{Avec-RM-spec}}
	\caption{(a) Probability densities $\mathcal{P}_{n}^{(\pm)}$ related to the bispinors $\mathbf{\Psi}_{1}^{(+)}$ (upper-blue), $\mathbf{\Psi}_{0}^{(-)}$ (lower-blue), and $\mathbf{\Psi}_{1}^{(-)}$ (lower-red). (b) Probability densities $\mathcal{P}_{n}^{(\pm)}$ related to the bispinors $\mathbf{\Xi}_{1}^{(+)}$ (upper-blue), $\mathbf{\Xi}_{0}^{(-)}$ (lower-blue), $\mathbf{\Xi}_{1}^{(-)}$ (lower-red). (c) Energy levels $\varepsilon^{(+)}_{n}$ (blue-solid), $\varepsilon^{(-)}_{n}$ (blue-dashed), $\widetilde{\varepsilon}^{(+)}_{n}$ (red-solid), and $\widetilde{\varepsilon}^{(-)}_{n}$ (red-dashed). In all the cases, the parameters have been fixed as $U_{0}=1$, $\kappa=2.5$, $\ell=1.1$, $\epsilon_1=-1$, $\epsilon_2=1.5$, $v_{3}=0.3$, and $\gamma=0.7$.}
	\label{SpRM}
\end{figure}

The bispinor solutions of (\ref{stac}) can be obtained via (\ref{PsiXi}). In Fig.~\ref{SpRM}, we illustrate calculated energy levels of $h_1$, $h_2$ and $H$, and probability density related to the bispinors $\mathbf{\Psi}_{n}$ and $\mathbf{\Xi}_{n}$. In particular, we have considered $U_{0}=1$, $\kappa=2.5$, $\ell=1.1$, $\epsilon_1=-1$, $\epsilon_2=1.5$, $v_{3}=0.3$, and $\gamma=0.7$. In such a case, we obtain $n_{max}=1$, so that we generate two physical solutions for each reduced Hamiltonian. On the other hand, for both $h_{1}$ and $h_{2}$, we get $V_{1}<V_{2}$, which means that $E_{0}^{(+)}$ is discarded from the point spectrum. Each reduced Hamiltonian contributes with three physical energies, and so the bilayer Hamiltonian $H$ contains six energy levels. See Fig.~\ref{Avec-RM-spec}.


\section{Confinement by the on-site interactions\label{four}}
\label{On-site-inter}
In this section, we focus on the case where $A(x)$ in (\ref{redHblg}) vanishes. The on-site interactions $\epsilon_1(x)$, $\epsilon_2(x)$ as well as the inter-layer coupling $\gamma(x)$ can be inhomogeneous. Let us suppose that $\epsilon_{2}(x)$ and $\gamma(x)$ are related by (\ref{gamma}). It brings the equations (\ref{eq1}) and (\ref{eq2}) into
\begin{align}
h_1\xi&=\left(-i
\begin{pmatrix}
v_{3} & 1 \\
1 & 0
\end{pmatrix}
\partial_{x}+
\begin{pmatrix}
\epsilon_1(x) &0 \\
0 & V_{2}
\end{pmatrix}\right)\xi=E\,\xi
\, ,\quad V_2=\epsilon_2+\gamma, \label{4eqa}
\\
h_2\chi&=\left(-i
\begin{pmatrix}
-v_{3} & 1 \\
1 & 0
\end{pmatrix}
\partial_{x}+
\begin{pmatrix}
\epsilon_1(x) &0 \\
0 & \epsilon_2(x)-\gamma(x)
\end{pmatrix}
\,\right)\chi=E\chi . \label{4eqb}
\end{align}
As discussed in the Section \ref{sectiontwo}, the solutions of (\ref{4eqa}) can be found via the Schr\"odinger equation with energy-dependent potential (\ref{psi1t}). It acquires the following simple form
\begin{equation}
-\widetilde{\psi}_{1}''-(E-V_{2})(E-\epsilon_{1}(x)) \widetilde{\psi}_{1} = \frac{v_{3}^{2}}{4}(E-V_{2})^{2} \widetilde{\psi}_{1} \vert_{V_2=\epsilon_2+\gamma}\, ,
\label{4psi1t}
\end{equation}
where the spinor components of $\xi=(\xi_1,\xi_2)^T$ are determined from $\widetilde{\psi}_{1}$ through
\begin{equation}
\xi_{1}=e^{-i\frac{v_3}{2}(E-V_{2})x}\widetilde{\psi}_{1} \, , \quad \xi_{2}=-\frac{i\,\psi_{1}'}{(V_{2}-E)} \, ,
\label{spin-comp}
\end{equation}
see (\ref{hom-spin1}) and (\ref{hom-spin2}).
The equation (\ref{4eqb}) reduces into (\ref{psi1t0}). With current fixing of the quantitites and denoting $\mathcal{B}(x)=\epsilon_2(x)-\gamma(x)$, the later equation reads as
\begin{equation}
-\widetilde{\chi}_{1}''+\left(-\frac{v_3^2}{4}(E-\mathcal{B}(x))^2
-(E-\epsilon_1(x))(E-\mathcal{B}(x))+\frac{3(\mathcal{B}(x)'(x))^2}{4(E-\mathcal{B}(x))^2}+\frac{\mathcal{B}(x)''(x)}{2(E-\mathcal{B}(x))}\right)\widetilde{\chi}_1=0 \, ,
\label{4psi1t0}
\end{equation}
where $\chi_1=\sqrt{E-\mathcal{B}(x)}e^{-i\frac{v_3}{2}\left(x-\int^x \mathcal{B}(s)ds\right)}\widetilde{\chi}_1$, see (\ref{pahse1}) and (\ref{hom-spin2}).

We find it physically reasonable to consider the systems where $\epsilon_1(x)$ is bounded.   We shall match either (\ref{4psi1t}) or (\ref{4psi1t0}) with the stationary equation of a solvable quantum systems that meets these requirements. Let us consider the stationary equation of the P\"oschl-Teller system
\begin{equation}\label{poschlteller}
-\widetilde{\psi}_1''-\kappa(\kappa-1)U_0^2\mbox{sech}^2U_0x\,\widetilde{\psi}_1-\lambda\widetilde{\psi}_1=0,\quad \kappa>1.
\end{equation}
It is worth noticing that Dirac electrons in bilayer graphene were studied in presence of P\"oschl-Teller electrostatic potential in \cite{Park}, see also \cite{Portnoi2}.
The equation (\ref{poschlteller}) is a special case ($\ell=1$) of the Rosen-Morse equation (\ref{pot-RM}) that was discussed in the previous section. 
Therefore, we can use (\ref{RMboundstates}) and write down the square integrable solutions $\widetilde{\psi}_{1;n}$ of (\ref{poschlteller}), 
\begin{equation}
\widetilde{\psi}_{1;n}=\mathcal{C}_{n}^{(\pm)}\operatorname{sech}(U_0 x)^{\alpha_n} P_{n}^{(\alpha_{n},\alpha_{n})}(\operatorname{tanh}(U_{0}x)) \, ,\label{4psi}
\end{equation}
where $n=0,1,\ldots,n_{max}=\lfloor \kappa-1 \rfloor$, and $\alpha_{n}= (\kappa-n-1).$ The corresponding eigenvalues $\lambda_n$ are
\begin{equation}\label{PT-lambda}
\lambda_n=-U_0^2(1-\kappa+n)^2=-U_0^2\,\alpha_n^2.
\end{equation}
Now, we shall identify either (\ref{4psi1t}) or (\ref{4psi1t0}) with (\ref{poschlteller}). Let us start with (\ref{4psi1t}),
\begin{equation}\label{identification2}
(E-V_{2})(E-\epsilon_{1}(x))  + \frac{v_{3}^{2}}{4}(E-V_{2})^{2}=\kappa(\kappa-1)\,U_0^2\,\mbox{sech}^2U_0x-U_0^2\,(1-\kappa+n)^2.
\end{equation}
There are different ways how to fix $\epsilon_1(x)$, and each of them leads to different values of $E$. Let us discuss some of them.

\subsection{Case I}
To begin with, let us consider the inhomogeneous on-site interaction
\begin{align}
\epsilon_1(x)=\mathcal{A}\,U_0^2\operatorname{sech}^2U_0x,\quad \mathcal{A}>0 . 
\label{id2}
\end{align}
Then (\ref{identification2}) is satisfied provided that we fix $\kappa_\epsilon\equiv\kappa$ and $E$ such that they solve the following two equations, 
\begin{align} 
\kappa_{\epsilon}(\kappa_{\epsilon}-1)=\mathcal{A}\,(V_2-E) \, , \quad 
U_0^2\,(1-\kappa_{\epsilon}+n)^2=(V_{2}-E)\left(\frac{v_{3}^{2}}{4}(V_{2}-E)-E\right)\, .\label{eq51}
\end{align}
That is, the P\"oschl-Teller amplitude $\kappa_{\epsilon}$ depends explicitly on the energy through
\begin{equation}
\kappa_{\epsilon}=\frac{1}{2}+\sqrt{\mathcal{A}(V_2-E)+\frac{1}{4}} \, .
\label{kappa}
\end{equation}
On the other hand, the second equation in (\ref{eq51}) yields to a fourth-order polynomial equation for $E$ of the form
\begin{equation}
(V_{2}-E)\left(\left(1+\frac{v_{3}^{2}}{4}\right)(V_{2}-E)-V_{2} \right)=-U_{0}^{2}\left(n+\frac{1}{2}-\sqrt{\mathcal{A}\,(V_{2}-E)+\frac{1}{4}} \right)^{2} \, ,
\label{Econd1}
\end{equation}
the solutions of which become unfeasible to obtain in the general setup. Despite such complexity, we can proceed further and obtain some additional information. 

The straightforward calculations show that the square-integrable condition~\eqref{PT-lambda} still holds in this case, with $\kappa\rightarrow\kappa_{\epsilon}$ given in~\eqref{kappa}. Thus, with the current choice of parameters, the solutions of (\ref{poschlteller}) 
\begin{align}
&\widetilde{\psi}_{1;n}=\mathcal{C}_{n}\operatorname{sech}(U_{0}x)^{\overline{\alpha}_{n}}P_{n}^{(\overline{\alpha}_{n},\overline{\alpha}_{n})}(\tanh(U_{0}x)) \, , 
\label{onsite-psi1}\\ 
&\overline{\alpha}_{n}=-n-1+\kappa_{\epsilon}=-n-\frac{1}{2}+\sqrt{\mathcal{A}(V_{2}-E)+\frac{1}{4}} \, ,
\end{align}
are square-integrable provided that $\overline{\alpha}_{n}$ is positive. In order to keep $\overline{\alpha}_n$ real, the term inside the square-root of~$\overline{\alpha}_{n}$ must be positive.  Since $\mathcal{A}>0$, we get $E< V_{2}+\frac{1}{4\mathcal{A}}$. Additionally, the requirement $\overline{\alpha}_n>0$ is sastified provided that $E< V_{2}-\frac{n(n+1)}{\mathcal{A}}$.  Now, the right-hand side of (\ref{Econd1}) is negative, as it a multiple of $-\overline{\alpha}_n^2$. Therefore,
one obtains real solutions of (\ref{Econd1}) for $(V_{2}-E)$ only if the term on the left is negative as well, which is quadratic and convex on $(V_{2}-E)$. We obtain $0<(V_{2}-E)<\frac{4V_{2}}{4+v_{3}^{2}}$, for $V_{2}>0$, and $\frac{4 V_{2}}{4+v_{3}^{2}}<(V_{2}-E)<0$, for $V_{2}<0$. Therefore, any real solution of~(\ref{Econd1}) has to lie inside the one of the following intervals
\begin{equation}
\begin{aligned}
&E\in\left(\frac{v_{3}^{2}}{4+v_{3}^{2}}V_{2},V_{2} \right)
\cap\left(-\infty,V_2-\frac{n(n+1)}{\mathcal{A}}\right) \, , && \operatorname{for} \, V_{2}>0 ,
\\& E\in\left(V_{2},\frac{v_{3}^{2}}{4+v_{3}^{2}}V_{2} \right)
\cap\left(-\infty,V_2-\frac{n(n+1)}{\mathcal{A}}\right) \, , \quad && \operatorname{for} \, V_{2}<0, 
\end{aligned}
\, 
\label{gap}
\end{equation}
where in the latter is clear that $\frac{v_{3}^{2}}{4+v_{3}^{2}}<1$, for $v_{3}\in\mathbb{R}$. The requirement that the intersections are non-empty sets an upper bound for possible values of $n$. Indeed, when $V_2>0$, the intersection is nonempty for $n\leq n_{max}$ where
\begin{equation}
n_{max}=\left\lfloor \sqrt{\frac{4\mathcal{A}V_{2}}{4+v_{3}^{2}}+\frac{1}{4}}-\frac{1}{2} \right\rfloor \, , \quad  \operatorname{for} \, V_{2}>0 \, .
\label{nmax}
\end{equation}
It provides us with an upper bound for the maximum number of physical solutions which is $n_{max}+1$. It is worth noticing that the trigonal warping term acts against the confinement here; the larger is $|v_3|$, the smaller is $n_{max}$.
When $V_{2}<0$, it is clear that only $n=0$ leads to an non-empty intersection of the energy intervals. This yields to $E=V_{2}$. Nevertheless, the expression (\ref{V2=E}) suggests that the corresponding spinor is not square integrable. Thus, such a solution is discarded, and no physical solutions are produced for $V_{2}<0$.

Interestingly, even if $E$ has to be found by numerical means, we have been able to extract general information about the spectrum and number of physically allowed solutions. Furthermore, the spinor may be computed explicitly from~\eqref{spin-comp} and~\eqref{onsite-psi1} as
\begin{multline}
{\psi}_{n}=
\begin{pmatrix}
\psi_{1;n} \\
\psi_{2;n}
\end{pmatrix}=\mathcal{N}_{n}\,e^{i\frac{v_{3}}{2}(V_{2}-E_{n})x}\left( \operatorname{sech}(U_{0}x)\right)^{\overline{\alpha}_{n}}\times\\
\begin{pmatrix}
P_{n}^{(\overline{\alpha}_{n},\overline{\alpha}_{n})}(z(x)) \\
-\left(\frac{v_{3}}{2}+i\frac{U_{0}\overline{\alpha}_{n}}{V_{2}-E_{n}}\tanh(U_{0}x)\right)P_{n}^{(\overline{\alpha}_{n},\overline{\alpha}_{n})}(z(x))+i\frac{U_{0}(n+1+2\overline{\alpha}_{n})}{2(V_{2}-E_{n})}\operatorname{sech}^{2}(U_{0}x)P_{n-1}^{(\overline{\alpha}_{n}+1,\overline{\alpha}_{n}+1)}(z(x))
\end{pmatrix} 
\, ,
\label{nh-xi}
\end{multline}
where $z(x):=\tanh(U_{0}x)$, and $n=0,1,\ldots,n_{max}$. The corresponding set of bispinors follow from~\eqref{ansatz}, which in this case are given through
\begin{equation}
\mathbf{\Psi}_{n}=\frac{1}{\sqrt{2}}(\psi_{1;n},\psi_{2;n},\psi_{2;n},\psi_{1;n})^{T} \, , \quad n=0,1,\ldots,n_{max} \, .
\end{equation}

\begin{table}
	\centering
	\begin{tabular}{c | c | c | c | c |} 
		\cline{2-5}
		{} & \multicolumn{2}{|c|}{$h_{1}$} & \multicolumn{2}{|c|}{$h_{2}$} \\
		\hline
		\multicolumn{1}{|c|}{$n$} & $E_{n}$ & $\overline{\alpha}_{n}$ & $E_{n}$ & $\overline{\alpha}_{n}$ \\ [0.5ex] 
		\hline
		\multicolumn{1}{|c|}{0} 	& 2.3 		& 0			& 1.7			& 0 \\ 
		\multicolumn{1}{|c|}{ } 	& 1.42592  	& 1.11555	& 1.18507		& 0.780747 \\
		\hline
		\multicolumn{1}{|c|}{1}	& 2.11215 	& -0.629828	& 1.47578		& -0.575127 \\
		\multicolumn{1}{|c|}{ } 	& 0.36825	& 0.837884 	& 0.217073		& 0.562499 \\
		\hline
		\multicolumn{1}{|c|}{2}	& 1.61515 	& -1.05117	& 0.779102		& -0.845786 \\
		\multicolumn{1}{|c|}{ }	& 0.0063725	& 0.038266	& 0.0792946		& -0.349208 \\
		\hline
		\multicolumn{1}{|c|}{3}	& 0.462606 + 0.648775 i 	& -1.18608 $-$ 0.378511 i		& 0.0890298 + 1.03522 i			& -1.26595 $-$ 0.62557 i \\
		\multicolumn{1}{|c|}{ } & 0.462606 $-$ 0.648775 i	& -1.18608 + 0.378511 i			& 0.0890298 $-$ 1.03522 i		& -1.26595 + 0.62557 i \\
		\hline
	\end{tabular}
	\caption{Numerical solutions of the characteristic equation~\eqref{Econd1}. We have fixed the parameters as $\mathcal{A}=2.7$, $v_{3}=0.1$, $U_{0}=1$. For $h_{1}$ we used $V_{2}=\epsilon_{2}+\gamma=2.3$, whereas for $h_{2}$ we have $V_{2}=\epsilon_{2}-\gamma=1.7$.}
	\label{T1}
\end{table}

\begin{figure}
	\centering
	\subfloat[][$h_{1}$]{\includegraphics[width=0.4\textwidth]{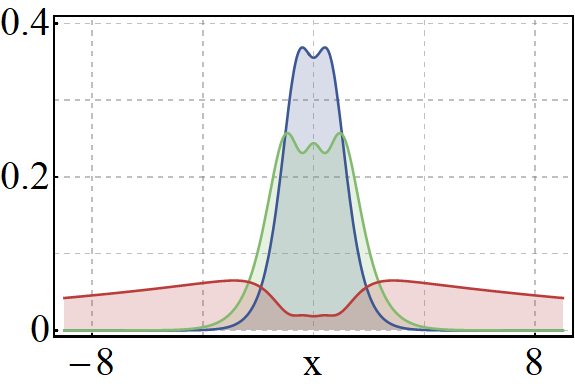}
		\label{FnonHom1a}}
	\hspace{2mm}
	\subfloat[][$h_{2}$]{\includegraphics[width=0.4\textwidth]{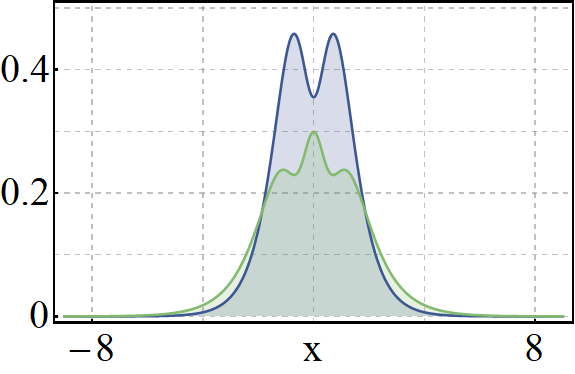}
		\label{FnonHom1b}}
	\caption{Normalized probability distributions $P_{n}=(\mathbf{\Psi}_{n}^{*\,})^{\,T}\cdot\mathbf{\Psi}_{n}$ (a) and $P_{n}=(\mathbf{\Xi}_{n}^{*\,})^{\,T}\cdot\mathbf{\Xi}_{n}$ (b) for $n=0$ (blue-solid), $n=1$ (green-solid), and $n=2$ (red-solid). In both cases, the parameters have been fixed as in Table~\ref{T1}.}
	\label{Fnonhom1}
\end{figure}

To illustrate our results, we consider $V_{2}=2.3$, $v_{3}=0.1$, $U_{0}=1$, and $\mathcal{A}=2.7$, so that real energies lie inside the interval $E\in(0.005735,2.3)$. Moreover, from~\eqref{nmax}, the maximum number of physical energies is $n_{max}+1=3$. The numerical values for $E_{n}$ and $\overline{\alpha}_{n}$ are shown in Table~\ref{T1}. For each $n$, we obtain two energies $E_{n}$, which are all real for $n=0,1,2$, and complex for $n\geq 3$. The physical energies are identified as $E_{0}=1.42592$, $E_{1}=0.36825$, and $E_{2}=0.0063725$ as they render $\overline{\alpha}_n$ positive. The associated probability distributions are depicted in Fig.~\ref{FnonHom1a}. 

\subsubsection*{Particular setup for $h_{2}$}
As we have remarked, the solutions of (\ref{4psi1t}) are insensitive with respect to the explicit form of $\gamma$ and $\epsilon_2$, and thus we can impose some further restrictions in order to obtain information about the reduced Hamiltonian $h_{2}$. Particularly, if 
\begin{equation}
\epsilon_2,\ \gamma\in\mathbb{R},
\end{equation}
the equation (\ref{4psi1t0}) brings us back to the system (\ref{eq51}), where now $V_2\equiv \mathcal{B}=\epsilon_2-\gamma$. Therefore, we can solve the equation in exactly the same manner as we did in Tab.\ref{T1} for different values of $V_2$ now. This is illustrated in the fourth and fifth columns of Table~\ref{T1}, where the energy levels are determined for $V_2=\epsilon_{2}-\gamma=1.6$, $v_3=0.1$, $U_{0}=1$, and $\mathcal{A}=2.7$. The corresponding probability distributions associated to the bispinors $\mathbf{\Xi}_{n}$ are depicted in Fig.~\ref{FnonHom1b}.


\subsection{Case II}
Fixing of $\epsilon_1(x)$ in (\ref{id2}) allowed us to have the on-site interaction independent on $n$. The price we paid was that the equation (\ref{Econd1}) had to be solved numerically. 
Let us consider the other choice of parameters such that (\ref{identification2}) is satisfied. We fix
\begin{equation}\label{id3}
\epsilon_1(x)=\kappa\, \mathcal{A}\,U_0^2\operatorname{sech}^2U_0x.
\end{equation}
Additionally, there must hold
\begin{equation}\label{eq57b}
-\mathcal{A}(E-V_2)=(\kappa-1)U_0^2,\quad
(E-V_{2})\left(\frac{v_{3}^{2}}{4}(E-V_{2})+E\right)=-U_0(1-\kappa+n)^2.
\end{equation}
We can see that inclusion of $\kappa$ into $\epsilon_1(x)$ lowered the order of $\kappa$ in the second equation in (\ref{eq57b}) when compared to (\ref{eq51}). We can solve (\ref{eq57b}) for $E$ and either $\kappa$, $U_0$ or $V_2$. As we require $\epsilon_1(x)$ to be independent of $n$, we solve the equation for $V_2$, and $E$,
\begin{align}
V_2(\mathcal{A},\kappa,n)&=\left(1+\frac{v_3^2}{4}\right)\frac{(\kappa-1)}{\mathcal{A}}+\frac{\mathcal{A}\,(1+n-\kappa)^2U_0^2}{\kappa-1},\nonumber\\
E(\mathcal{A},\kappa,n)&=\frac{\mathcal{A}\,U_0^2\,(1+n-\kappa)^2}{\kappa-1}+\frac{(\kappa-1)v_3^2}{4\mathcal{A}}.\label{V2E}
\end{align}
$V_2(\mathcal{A},\kappa,n)$ is a parabola in $n$ with minimum $V_2(n_0)=\frac{(\kappa-1)}{\mathcal{A}}\left(1+\frac{v_3^2}{4}\right)$ at $n_0=\kappa-1$. 
For each $n$ from the allowed interval $n\in\{0,\dots,\lfloor\kappa-1\rfloor\}$, the on-site interaction $\epsilon_1(x)$ remains the same. Nevertheless, the value of $V_2$ gets changed correspondingly.  

\begin{figure}
	\centering
	\includegraphics[width=0.4\textwidth]{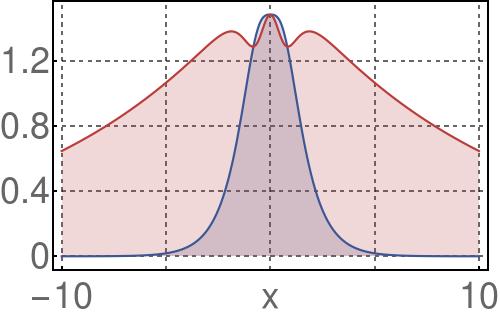} 
	\caption{Density of probability of $\mathbf{\Psi}$ and $\mathbf{\Xi}$ from (\ref{caseIIPsi}) and (\ref{caseIIXi}), respectively, for $n=0$, and $\widetilde{n}=1$, $\kappa=2.1$, $\mathcal{A}=1$, $U_0=1$, $v_3=0.1$. }
	\label{caseII}
\end{figure}

For each of this specific configurations, we are able to find a localized solution $\widetilde{\psi}_{1;n}$. The bispinor solution of (\ref{stac}) is then
\begin{equation}
\mathbf{\Psi}=\left(\psi_1,\frac{-i\psi_1'}{E-V_2(\mathcal{A},\kappa,n)},\frac{-i\psi_1'}{E-V_2(\mathcal{A},\kappa,n)},\psi_1\right),\quad \psi_1=e^{-i\frac{v_3}{2}(E-V_2)x}\widetilde{\psi}_{1;n}\vert_{\substack{E=E(\mathcal{A},\kappa,n)\\V_{2}=V_2(\mathcal{A},\kappa,n)}}.
\end{equation}
The solution is invariant with respect to the changes of $\epsilon_2(x)$ and $\gamma(x)$ that preserve $V_2$, including the case where both $\epsilon_2(x)$ and $\gamma(x)$ are constant. When this is the case, the equation (\ref{4psi1t0}) reduces into the equation that coincides with (\ref{4psi1t}), yet for $V_2=\epsilon_2-\gamma$. If $\epsilon_2$ and $\gamma$ are such that 
\begin{equation}
\epsilon_2+\gamma=V_2(\mathcal{A},\kappa,n)\quad \mbox{and}\quad \epsilon_2-\gamma=V_2(\mathcal{A},\kappa,\widetilde{n}),\quad n,\widetilde{n}\in\{0,\dots,\lfloor\kappa-1\rfloor\},\label{caseIIPsi}
\end{equation}
then we can get bound state solutions  for each equation (\ref{4eqa}) and (\ref{4eqb}) with energies $\epsilon=E(\mathcal{A},\kappa,n)$ and $\widetilde{\varepsilon}=E(\mathcal{A},\kappa,\widetilde{n})$. 
The bispinor $\mathbf{\Xi}$ corresponding to the later energy is  is
\begin{equation} 
\mathbf{\Xi}=\left(\psi_1,\frac{-i\psi_1'}{E-V_2(\mathcal{A},\kappa,\widetilde{n})},\frac{i\psi_1'}{E-V_2(\mathcal{A},\kappa,\widetilde{n})},-\psi_1\right),\quad \psi_1=e^{i\frac{v_3}{2}(E-V_2)x}\widetilde{\psi}_{1;\widetilde{n}}\vert_{\substack{E=E(\mathcal{A},\kappa,\widetilde{n})\\V_{2}=V_2(\mathcal{A},\kappa,\widetilde{n})}}\label{caseIIXi}.
\end{equation}
We show density of states of the corresponding bispinors $\mathbf{\Psi}$ and $\mathbf{\Xi}$ in Fig.\ref{caseII}.

\subsection{Case III}
Let us relax the condition (\ref{gamma}), i.e. both $\epsilon_2+\gamma$ and $\epsilon_2-\gamma$ can be inhomogeneous. In this case,  the equations (\ref{eq1}) and (\ref{eq2}) lead to (\ref{psi1t0}), yet with different form of $V_2(x)$ in each case. As we mentioned in section \ref{sectiontwo}, we can find configuration of $\epsilon_1(x)$, $\epsilon_2(x)$ and $\gamma(x)$ such that (\ref{psi1t0}) is partially solvable. Let us consider the case with $V_2(x)=\gamma(x)+\epsilon_2(x)$. We fix $V_1(x)\equiv \epsilon_1(x)$ where
\begin{equation}\label{V1}
\epsilon_1(V_2,V_0,E)=E+\frac{v_3^2}{4}(E-V_2(x))-\frac{3(V_2'(x))^2}{4(E-V_2(x))^3}-\frac{V_2''(x)}{2(E-V_2(x))^2}+\frac{V_0(x)}{E-V_2(x)}.
\end{equation}
Then the equation (\ref{psi1t0}) reduces into 
\begin{equation}\label{rcesolvable}
-\widetilde{\psi}_1''+V_0(x)\widetilde{\psi}_1=0.
\end{equation}
Let us identify (\ref{rcesolvable}) with the stationary equation of the P\"oschl-Teller system again, $V_0(x)=-\kappa(\kappa-1)U_0^2\operatorname{sech}^2U_0x+U_0^2(1+n-\kappa)^2$. When $n$ is a positive integer $n\in\{0,\lfloor \kappa-1\rfloor\}$, the equation (\ref{rcesolvable}) has a localized solution $\widetilde{\psi}_{1;n}$, see (\ref{4psi}). Notice that this solution is independent on the explicit choice of $V_2(x)$ and $E$ as they do not appear in (\ref{rcesolvable}). Nevertheless, both $V_2(x)$ and $E$ affect the form of the bispinor solution (\ref{PsiXi}) via (\ref{pahse1}),
\begin{equation} 
\mathbf{\Psi}=\left(\psi_1,\frac{-i\psi_1'}{E-\epsilon_2-\gamma},\frac{-i\psi_1'}{E-\epsilon_2-\gamma},\psi_1\right),\quad \psi_1=\sqrt{E-\epsilon_2-\gamma}e^{-i\frac{v_3}{2}(E-\epsilon_2-\gamma)x}\widetilde{\psi}_{1;\widetilde{n}}
\label{caseIII}.
\end{equation}
In the current setting, $E$ plays rather the role of an interaction parameter.  We can tune the interaction $\epsilon_1(x)$ by changing $E$ such that it confines a bound state with energy equal to $E$. In order to keep $\mathbf{\Psi}$ square-integrable, we require that $\sqrt{E-V_2(x)}$ is a bounded function.  

\begin{figure}
	\centering
	\includegraphics[width=0.4\textwidth]{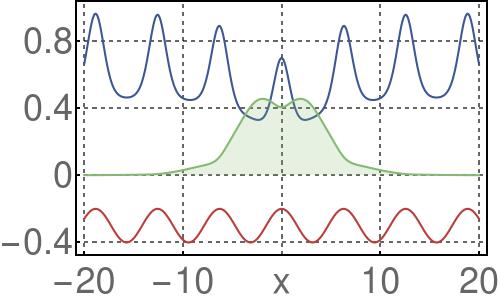}   \includegraphics[width=0.4\textwidth]{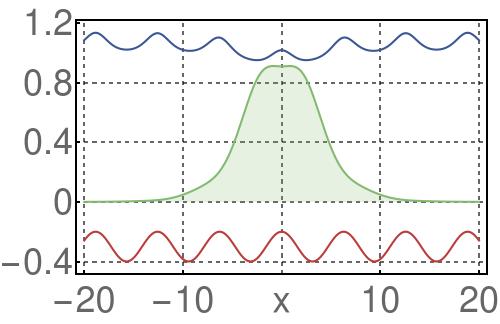}  
	\caption{$\epsilon_1(x)$ from (\ref{epsilon1}) is blue, $\epsilon_2$ from (\ref{epsilon2}) is red, $\Psi^\dagger\Psi$ is green. We fixed $\kappa=2.2$, $v_3=0.2$, $U_0=0.2$, $c=0.1$, $n=0$, $\gamma=0.5$, and $E=0.5$ (left) and $E=1$ (right).}
	\label{SpOsc3}
\end{figure}

We fix $V_2(x)=\epsilon_2(x)+\gamma$, $\gamma\in\mathbb{R}$. When $\epsilon_2(x)$ is periodic, $\epsilon_1(x)$ shares its periodicity up to the last term in (\ref{V1}) that represents a periodicity defect. For instance, if we fix
\begin{align}\label{epsilon2}
\epsilon_2(x)=&V_2(x)-\gamma=c \cos x,\quad\quad 0<c<E,
\end{align}
then $\epsilon_1(x)$ reads explicitly
\begin{align}
\epsilon_1(x)=&E+\frac{(E-c \cos x) v_3^2}{4}+\frac{-2 c \cos x(c \cos x -E)-3c^2\sin^2x}{4(E-c \cos x)^3}\\&+\frac{(1+n-\kappa)^2U_0^2}{E-c\cos x}+\frac{(1-\kappa)\kappa U_0^2 \operatorname{sech}^2U_0 x }{E-c\cos x}.
\label{epsilon1}\end{align}
We illustrate the interactions for different values of parameters in Fig.~\ref{SpOsc3} together with density of probability of the bound state.

\section{Confinement by the inter-layer interaction}
\label{five}
Up to now, the inter-layer coupling $\gamma$ had rather implicit influence on considered solutions as it was "hidden" in $V_2$. Let us see whether we can get an analytical solution of confined Dirac fermions by inhomogeneous $\gamma$. It is worth noticing in this context that confinement by inhomogeneous $\gamma$ with rotational symmetry was analyzed numerically in \cite{Abdullah}, \cite{Solomon}. We focus on the situation where $\epsilon_1$ and $\epsilon_2$ are constant (in order to eliminate localization by on-site potentials) and $\gamma$ is inhomogeneous. We fix
\begin{equation}
v_3=0,\quad V_1=\epsilon_1=const., \quad \epsilon_2=const., \quad V_2(x)=\epsilon_2+\gamma(x).
\end{equation}
Then the equations (\ref{eq1})-\eqref{eq2} for the spinor components $\xi_j$ and $\chi_j$, for $j=1,2$, can be decoupled through the relationships
\begin{equation}
\label{4coupled2}
-\frac{i\xi_{2}'}{E-\epsilon_1}=\xi_1, \quad -\frac{i\chi_{2}'}{E-\epsilon_1}=\chi_1 \, ,
\end{equation}
leading to the effective energy-dependent Schr\"odinger equation
\begin{equation}
\label{schr-layer}
-\xi_{2}''-(\epsilon_1-E)\gamma(x)\xi_{2}=(\epsilon_{1}-E)(\epsilon_{2}-E)\xi_{2} \, , \quad -\chi_{2}''+(\epsilon_1-E)\gamma(x)\chi_{2}=(\epsilon_{1}-E)(\epsilon_{2}-E)\chi_{2} \, .
\end{equation}
Notice that the difference between the latter equations relies on the sign of the energy-dependent potential, which both coincide qualitatively with (\ref{4psi1t}), therefore, we can follow the same steps as in the previous section. We shall identify both equations in~\eqref{schr-layer} with (\ref{poschlteller}). We set\footnote{We have set, without loss of generality, $\mathcal{A}>0$ to simplify the conditions for the reality of the spectrum. Nevertheless, similar conclusions can be withdrawn if we allow $\mathcal{A}<0$.}
\begin{equation}\label{id4}
\gamma(x)=\mathcal{A}\, U_0^2\operatorname{sech}^2 U_0x +\gamma_0\, , \quad \gamma_0,\,\mathcal{A}> 0 \, , \quad U_0\in\mathbb{R} \, .
\end{equation}
It has similar form to $\epsilon_1(x)$ in (\ref{id2}), however, it acquires nonvanishing constant value $\gamma_0$ asymptotically now. 
In analogy to the previous section, we identify the set of solutions as
\begin{equation}
\begin{aligned}
& \psi_{2;n}=\mathcal{C}_{2}(\operatorname{sech}(U_{0}x))^{\eta_{n}} P_{n}^{(\eta_{n},\eta_{n})}(\tanh(U_{0}x)) \, , \\
& \chi_{2;n}=\mathcal{D}_{2}(\operatorname{sech}(U_{0}x))^{\widetilde{\eta}_{n}}P_{n}^{(\widetilde{\eta}_{n},\widetilde{\eta}_{n})}(\tanh(U_{0}x)) \, ,
\end{aligned}
\label{inter-sols1}
\end{equation}
where $\mathcal{C}_{2}$ and $\mathcal{D}_{2}$ are the respective normalization factors, and
\begin{equation}
\eta_{n}=\nu_{E}-n-1 \, , \quad \widetilde{\eta}_{n}=\widetilde{\nu}_{E}-n-1 \, , 
\label{eta}
\end{equation}
together with
\begin{equation}
\nu_{E}=\frac{1}{2}+\sqrt{\mathcal{A}(\epsilon_{1}-E)+\frac{1}{4}} \, , \quad 
\widetilde{\nu}_{E}=\frac{1}{2}+\sqrt{-\mathcal{A}(\epsilon_{1}-E)+\frac{1}{4}} \, .
\label{inter-kappa}
\end{equation}
On the other hand, from the relationship
\begin{equation}
(\epsilon_{1}-E)(\epsilon_{2}-E+\delta\,\gamma_0)=-U_{0}^{2}\left(n+\frac{1}{2}-\sqrt{\delta\mathcal{A}(\epsilon_{1}-E)+\frac{1}{4}} \right)^{2} \, , \quad \delta= +1,-1 \, ,
\label{inter-E1}
\end{equation}
we extract the energies of $h_{1}$ and $h_{2}$ after choosing $\delta=+1$ and $\delta=-1$, respectively. 

It is worth to recall that an immediate solution for the energy equation~\eqref{inter-E1} can be found for $n=0$ and $E=\epsilon_{1}$ in both cases $\delta=\pm 1$. Nevertheless, such a solution is discarded as it is not square-integrable.

Likewise in (\ref{gap}), we can obtain the energy intervals in which $E$ takes real values for $h_{1}$ and $h_{2}$. First, we should guarantee that $\eta_n$ and $\widetilde{\eta}_n$ in~\eqref{eta} are real and positive in order to get square-integrable solutions. Next, the left term in \eqref{inter-E1} should be negative as the equation would have no real solutions otherwise. 
Combining both results, we get the intervals where the real roots of (\ref{inter-E1}) have to lie,
\begin{eqnarray}
h_{1}: \, E\in
\left(\epsilon_{2}+\gamma_0,\epsilon_1-\frac{n(n+1)}{\mathcal{A}}\right),\quad \epsilon_1>\epsilon_2+\gamma_0,\\
h_{2}: \, E\in
\left(\epsilon_{2}+\gamma_0,\epsilon_1-\frac{n(n+1)}{\mathcal{A}}\right),\quad \epsilon_1<\epsilon_2-\gamma_0,
\label{inter-E2}
\end{eqnarray}
One cannot get square-integrable eigenstates of either $h_1$ or $h_2$ corresponding to real energies for other values of $\epsilon_1$. Now, the intervals (\ref{inter-E2}) are nonempty for some values of $n$ only. This way, we get an upper bounds $n_{max}$ and $\widetilde{n}_{max}$ for the number of bound states of $h_1$ and $h_2$, respectively, that we can obtain this way. They are  
\begin{equation}
n_{max}=\left\lfloor \sqrt{\mathcal{A}(\epsilon_{1}-\epsilon_{2}-\gamma_{0})+\frac{1}{4}}-\frac{1}{2} \right\rfloor \, , \quad \widetilde{n}_{max}=\left\lfloor \sqrt{\mathcal{A}(\epsilon_{2}-\epsilon_{1}-\gamma_{0})+\frac{1}{4}}-\frac{1}{2} \right\rfloor \, .
\label{inter-nmax}
\end{equation}
The corresponding bispinors are given by
\begin{equation}
\mathbf{\Psi}_{n}=\frac{1}{\sqrt{2}}\left(-i\frac{\psi'_{2;n}}{E-\epsilon_{1}},\psi_{2;n},\psi_{2;n},-i\frac{\psi'_{2;n}}{E-\epsilon_{1}} \right) \, , \quad 
\mathbf{\Xi}_{n}=\frac{1}{\sqrt{2}}\left(-i\frac{\xi'_{2;n}}{E-\epsilon_{1}},\xi_{2;n},-\xi_{2;n},i\frac{\xi'_{2;n}}{E-\epsilon_{1}} \right) \, ,
\label{inter-bispinor}
\end{equation}
for $\epsilon_{1}>\epsilon_{2}+\gamma_0$ and $\epsilon_{1}<\epsilon_{2}-\gamma_0$, respectively, with $\psi_{2;n}$ and $\chi_{2;n}$ given in~\eqref{inter-sols1}. 

\begin{figure}
	\centering
	\includegraphics[width=0.4\textwidth]{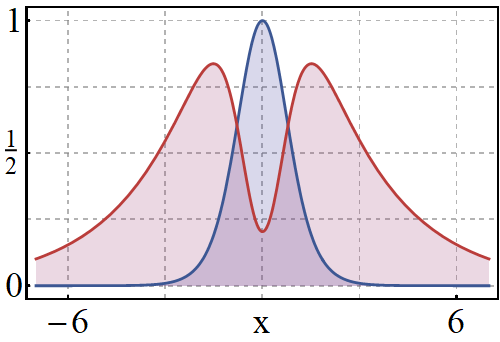}
	\caption{Probability density related to the bispinors $\mathbf{\Psi}_{n}$ of~\eqref{inter-bispinor} for $n=0$ (blue-solid) and $n=1$ (red-solid). The parameters has been fixed as in Table~\ref{inter-tab2}.}
	\label{inter-fig1}
\end{figure}

To illustrate the results presented in this section, let us fix the parameters as $U_{0}=\mathcal{A}=1$, $\epsilon_{1}=1.5$, $\epsilon_{2}=-1.5$, and $\gamma_{0}=0.3$. Since $\epsilon_{1}>\epsilon_{2}+\gamma_{0}$, we would expect bound states only for $h_{1}$, besides the non-physical solution $E=\epsilon_{1}=1.5$. Moreover, from~\eqref{inter-nmax}, one may see that only two bound states can be generated. Such an information may be verified in Table~\ref{inter-tab2}, where we obtain two physical energies for $h_{1}$ as $E_{0}=-0.685308$ and $E_{1}=-1.18315$. Although there are more real energies, they do not satisfy the finite-norm condition $\eta_{n}>0$. The behavior for the corresponding probability densities $\mathcal{P}_{n}$ is depicted in Fig.~\ref{inter-fig1}.

\begin{table}
	\centering
	\begin{tabular}{c|c|c|}
		\cline{2-3}
		{} & \multicolumn{2}{c|}{$h_{1}$} \\
		\hline
		\multicolumn{1}{|c|}{$n$} 	& $E$ 		& $\eta_{n}$ \\
		\hline
		\multicolumn{1}{|c|}{$0$} 	& -0.685308 	& 1.06055 	\\
		\multicolumn{1}{|c|}{}   	& 1.5 		& 0			\\
		\hline
		\multicolumn{1}{|c|}{$1$} 	& -1.18315	& 0.212643 	\\
		\multicolumn{1}{|c|}{}   	& 1.24511	& -0.789447 	\\
		\hline
		\multicolumn{1}{|c|}{$2$} 	& -0.873539 	& -0.880266 	\\
		\multicolumn{1}{|c|}{}  		& 0.381221	& -1.33005 	\\
		\hline
	\end{tabular}
	\caption{Energy solutions of~\eqref{inter-E1}, together with the finite-norm condition $\eta_{n}>0$ and $\widetilde{\eta}_{n}>0$, for the reduced Hamiltonians $h_{1}$ and $h_{2}$. The parameters have been fixed as $U_{0}=\mathcal{A}=1$, $\epsilon_{1}=1.5$, $\epsilon_{2}=-1.5$, and $\gamma_0=0.3$.}
	\label{inter-tab2}
\end{table}

\section{Discussion}
\label{conclu}
In the article, we focused on the systems described by Dirac Hamiltonians of the form~(\ref{redHblg}) that appear in the analysis of bilayer Dirac materials. We were interested in analytical treatment of confined states that can appear due to local fluctuations (\ref{id2}), (\ref{id3}), (\ref{id4}) or periodicity defects (\ref{epsilon1}) of the involved interactions.  

We have made use of the fact that the equation (\ref{stac}) is reducible in terms of the equations (\ref{eq1}) and (\ref{eq2}) with $2\times2$ Hamiltonians. 
In section \ref{sectiontwo}, we showed that solution of any of the two equations is equivalent to solution of Schr\"odinger equation (\ref{psi1t0}) whose potential is a nonlinear function of the interactions and their derivatives. We focused on the specific case where (\ref{psi1t0}) can be significantly simplified into Schr\"odinger equation with energy-dependent potential (\ref{psi1t}).

We considered confinement by a combination of external magnetic field and mechanical deformations in section \ref{three} where the energy-dependent Schr\"odinger equation was identified with the stationary equation of the harmonic oscillator or the Rosen-Morse system. 
In section \ref{four}, we focused on confinement by inhomogeneities of the on-site and inter-layer interactions. 
We showed that Dirac fermions can be confined by a local fluctuation or periodicity defect of the on-site interaction $\epsilon_1$. We demonstrated this fact on the systems with  P\"oschl-Teller-type interactions (\ref{id2}), (\ref{id3}), or periodic interactions with a localized defect (\ref{epsilon1}).  
Finally, we considered situation where only the interlayer interaction was inhomogeneous in section \ref{five}. Here we fixed the trigonal warping term vanishing. It allowed us analytical treatment of decoupled equations with energy-dependent potential (\ref{schr-layer}). 

In all the scenarios, we faced the need to solve Schr\"odinger equation with energy-dependent potential. In section \ref{three}, the later equation occured due to presence of the trigonal warping term, $v_3\neq0$. If the later term was absent, decoupling of (\ref{coupled1}) would produce Schr\"odinger equation with potential independent of energy. In sections $4$ and $5$, $\epsilon_1$, $\epsilon_2$ or $\gamma$ were inhomogeneous, so that there was an electrostatic component in the potential term of the reduced equations (\ref{eq1}) and (\ref{eq2}).  It is known \cite{Ghosh} that decoupling of components in the stationary equation for $2\times 2$ Dirac Hamiltonian with electric potential leads to Schr\"odinger equation with energy-dependent potential. The problems related to the solution of such an equation are avoided when bound states with zero energy are of interest, see e.g. \cite{Portnoi}, \cite{Portnoi3}, \cite{Ho} \cite{Schulze2}. The zero modes in presence of ihnomogeneous electric potential and an effective mass were discussed recently in \cite{Schulze3}.  The Hamiltonian (\ref{redHblg}) can be understood as two, coupled $2\times 2$ Dirac Hamiltonians. As we assumed that $\epsilon_1\neq\epsilon_2$ in general, there was electrostatic potential accompanied by an effective mass term. For these systems, we found localized states with energies distinct from zero.

\appendix

\section*{Acknowledgment}
M.C.-C. thanks Department of Physics of the Nuclear Physics Institute of CAS for hospitality.  M.C.-C. acknowledges the support of CONACYT, project FORDECYT-PRONACES/61533/2020. M.C.-C. also acknowledges the Conacyt fellowship 301117. V. J. was supported by GA\v CR grant no 19-07117S. K.Z. acknowledges the support from the project ``Physicist on the move II'' (KINE\'O II), Czech Republic, Grant No. CZ.02.2.69/0.0/0.0/18\_053/0017163.


\end{document}